\documentclass[aps,prl,twocolumn,superscriptaddress]{revtex4}
\usepackage{amsmath}
\usepackage{graphicx}
\usepackage{color}
\usepackage{longtable}
\usepackage{lineno}
\begin{document}


\title{Search for effects beyond the Born approximation in polarization transfer
observables in $\vec{e}p$ elastic scattering}


\author{M. Meziane}\email[Corresponding author: ]{mezianem@jlab.org}
\affiliation{The College of William and Mary, Williamsburg, Virginia 23187, USA}
\author{E. J. Brash}
\affiliation{Christopher Newport University, Newport News, Virginia 23606, USA}
\affiliation{Thomas Jefferson National Accelerator Facility, Newport News, Virginia 23606, USA}
\author{R. Gilman}
\affiliation{Rutgers, The State University of New Jersey,  Piscataway, New Jersey 08855, USA}
\affiliation{Thomas Jefferson National Accelerator Facility, Newport News, Virginia 23606, USA}
\author{M. K. Jones}
\affiliation{Thomas Jefferson National Accelerator Facility, Newport News, Virginia 23606, USA}
\author{W. Luo}
\affiliation{Lanzhou University,222 Tianshui Street S., Lanzhou 730000, Gansu, People's Republic of China}
\author{L. Pentchev}
\author{C. F. Perdrisat}
\affiliation{The College of William and Mary, Williamsburg, Virginia 23187, USA}
\author{A. J. R. Puckett}
\affiliation{Massachusetts Institute of Technology, Cambridge, Massachusetts 02139, USA}
\affiliation{Los Alamos National Laboratory, Los Alamos, New Mexico 87545, USA}
\author{V. Punjabi}
\author{F. R. Wesselmann}
\affiliation{Norfolk State University, Norfolk, Virginia 23504, USA}
% ...

\date{\today}

\begin{abstract}
Intensive theoretical and experimental efforts over the past decade have 
aimed at explaining the discrepancy between data for the proton electric 
to magnetic form factor ratio, $G_{E}/G_{M}$, obtained separately
from cross section and polarization transfer measurements. One 
possible explanation for this difference is a two-photon-exchange (TPEX)
contribution. In an effort to search for effects beyond
the one-photon-exchange or Born approximation, we report measurements of polarization transfer 
observables in the elastic $H(\vec{e},e'\vec{p})$ reaction for three different beam 
energies at a fixed squared momentum transfer $Q^2 = 2.5$ GeV$^2$, spanning a wide range 
of the virtual photon polarization parameter, $\epsilon$. 
From these measured polarization observables, we have obtained separately the ratio $R$, which equals $\mu_p G_{E}/G_{M}$ in the Born 
approximation, and the longitudinal polarization transfer component $P_\ell$, 
with statistical and systematic uncertainties of $\Delta R \approx \pm 0.01 
\mbox{(stat)} \pm 0.013 \mbox{(syst)}$ and $\Delta P_\ell/P^{Born}_{\ell} \approx \pm 0.006 
\mbox{(stat)}\pm 0.01 \mbox{(syst)}$. The ratio $R$ is found to be independent 
of $\epsilon$ at the 1.5\% level, while the $\epsilon$ dependence of $P_\ell$ shows 
an enhancement of $(2.3 \pm 0.6) \%$ relative to the Born approximation at large 
$\epsilon$. 
\end{abstract}
\maketitle
\paragraph{}
After decades of experimental and theoretical efforts, the internal 
structure of the nucleon remains one of the defining problems of nuclear 
physics. Based on the generally accepted notion that the 
electromagnetic interaction is well understood from a theoretical 
point of view, elastic electron-nucleon scattering has served as a 
powerful tool to measure fundamental observables: the 
electromagnetic form factors.  There are two experimental methods for 
extracting the ratio of the electric to magnetic form factors of the 
proton, $G_{E}/G_{M}$, from electron-proton elastic scattering. In the Rosenbluth separation technique \cite{Rosenbluth1950}, 
$G_E^2$ and $G_M^2$ are determined from the angular dependence of the reduced cross section $\sigma_{r}$ at constant $Q^2$.
Polarization experiments determine $G_E/G_M$ by using a polarized electron beam with either a polarized proton target, 
or a measurement of the transferred polarization to the scattered proton. At values of the squared-momentum-transfer, 
$Q^2 \le$ 1 GeV$^2$, data from the Rosenbluth and polarization techniques are in good agreement. 
At large $Q^2$, however, the cross section data \cite{PerdrisatPunjabiVanderhaegen2007,Andi94,
Christy04,Qattan05} disagree with the ratios $G_{E}/G_{M}$ obtained using the polarization transfer method 
\cite{Jones00,Punjabi05,Gayou02,Puckett10}. 
This systematic difference is a source of intense debate in both the  
theoretical and experimental nuclear physics communities. 

Recently, theoretical attention has 
been paid to the set of radiative corrections \cite{MoTsai,Walker94} that must be made to the cross section data in 
order to extract the form factor ratio. These corrections change the slope of the reduced 
cross section by as much as 30\% for larger $Q^2$. In contrast, radiative 
corrections to the polarization data are essentially negligible 
\cite{AfanasevRadCorr}. Until recently, only ``standard" radiative corrections 
were taken into account.  Based on the observed 
discrepancy in the data, new efforts have been made to include higher-order 
radiative mechanisms, such as two-photon exchange (TPEX) 
\cite{Afa05,Blund05,Kond05,Arri05,Byst07,CarlVdh07,Kivel09}. Several of these calculations 
have indicated that TPEX partially resolves the disagreement between 
the two data sets, but further investigation is needed. 
What is still lacking is a complete set of elastic $ep$ scattering 
observables sensitive to the TPEX amplitudes, with sufficient precision to 
guide the development of a consistent theoretical framework for the interpretation of 
the discrepancy in terms of TPEX. This experiment is an effort to provide additional data.

When considering the exchange of two
or more photons, the hadronic vertex function can generally be expressed, in terms of three independent and complex amplitudes, 
$\tilde{G}_{E,M}\equiv G_{E,M}(Q^2)+\delta\tilde{G}_{E,M}(Q^2,\epsilon)$ and $\tilde{F}_{3}$ which are functions 
of  $Q^2$ and the kinematical parameter $\epsilon=(1+2(1+\tau)\tan^{2}\theta_{e}/2)^{-1}$, where $\tau \equiv Q^{2}/4M^{2}$, $M$ 
is the proton mass,  and $\theta_{e}$ is the electron scattering
angle. In the Born approximation, the first two amplitudes equal the real electric and magnetic Sachs form factors 
which depend only on $Q^2$, and $\tilde{F}_{3}$ vanishes.
The experimental observables used to extract $G_{Ep}$ and $G_{Mp}$ from cross section and polarization transfer 
measurements, assuming the validity of the Born approximation, are affected in different ways by 
TPEX, as shown in Eq.(1). The Rosenbluth method relies on measuring the $\epsilon$ dependence of the 
reduced cross section $\sigma_r$ at fixed $Q^2$ to separate $G_E^2$ from $G_M^2$, and becomes highly sensitive 
to additive TPEX effects at large $Q^2$ when $\epsilon G_E^2/\tau G_M^2$ becomes small. The transferred 
polarization to the recoil proton in $H(\vec{e},e'\vec{p})$ has transverse ($P_t$) and longitudinal ($P_\ell$) 
components with respect to the momentum transfer in the scattering plane \cite{AkhiezerRekalo2,ArnoldCarlsonGross}. 
The ratio $R$ defined in Eq.(1) equals $\mu_p G_{E}/G_{M}$ in the Born approximation, and is much less 
vulnerable to TPEX corrections. The TPEX 
corrections appear in $\sigma_r$, $P_t$, $P_\ell$, and $R$ as interference terms between the Sachs form factors 
and the real part of the TPEX amplitudes \cite{Vander03}:
\begin{eqnarray}
P_{t}&=&-\frac{hP_{e}}{\sigma_{r}}\sqrt{\frac{2\epsilon(1-\epsilon)}{\tau}}\left[G_{E}G_{M}+G_{M}  \Re{\left(\delta\tilde{G}_{E}+\frac{\nu}{M^{2}}\tilde{F}_{3}\right)} \right.\nonumber\\
& &\left.+G_{E} \Re{(\delta \tilde{G}_{M})} +\mathcal{O}(e^{4})\right]\nonumber\\
P_{\ell}&=&\frac{hP_{e}}{\sigma_{r}}\sqrt{1-\epsilon^{2}}\left[G^{2}_{M}+2G_{M}\Re{\left(\delta \tilde{G}_{M}+\frac{\nu}{M^{2}}\frac{\epsilon}{1+\epsilon}\tilde{F}_{3}\right)}\right. \nonumber\\
& &\left.+\;\mathcal{O}(e^{4})\right]\nonumber\\
\sigma_{r}&=&G_{M}^{2}+\frac{\epsilon}{\tau}G_{E}^{2}+\frac{2\epsilon}{\tau}{G_{E}}\Re{\left(\delta\tilde{G}_{E}+\frac{\nu}{M^{2}}\tilde{F}_{3}\right)}\nonumber\\
& &+\;2G_{M}{\Re{\left(\delta\tilde{G}_{M}+\frac{\epsilon\nu}{M^{2}}\tilde{F}_{3}\right)}}+\mathcal{O}(e^{4}) \nonumber\\
R&\equiv&-\mu_{p}\sqrt{\frac{(1+\epsilon)\tau}{2\epsilon}}\frac{P_{t}}{P_{\ell}}=\mu_{p}\frac{G_{E}}{G_{M}}\Re\left[1-\frac{\delta\tilde{G}_{M}}{G_{M}}+\frac{\delta \tilde{G}_{E}} {G_{E}}\right. \nonumber\\
& &\left. +\frac{\nu\tilde{F}_{3}}{M^{2}}\left(\frac{1}{G_{E}}-\frac{2\epsilon}{1+\epsilon}\frac{1}{G_{M}}\right)\right]+\mathcal{O}(e^{4})\label{eq:twogammapol}
\end{eqnarray}
where $\Re$ stands for the real part, $h=\pm 1$ and $P_e$ are the helicity and polarization of the electron beam, 
and $\nu = M\frac{E_e+E'_e}{2}$, with $E_e$, $E'_e$ being the energy of the incident and scattered electron, 
respectively. In the Born approximation, these corrections vanish and the well known
expressions for these observables \cite{AkhiezerRekalo2,ArnoldCarlsonGross} are
recovered. Other observables, such as the induced normal polarization component, and the target-normal and beam-normal single-spin 
asymmetries, depend only on the imaginary (absorptive) part of the
TPEX amplitude. A direct test of the TPEX effect is the comparison between $e^{+}p$ and $e^{-}p$ elastic
scattering cross sections. Since the two-photon contributions
(relative to the Born amplitudes) are of opposite sign, a few percent
deviation from unity as a function of $\epsilon$ is predicted for the
ratio $\sigma _{e^{+}}/\sigma _{e^{-}}$. Recent analyses of $e^{\pm}p$ cross sections are inconclusive due to
large uncertainties in the data \cite{Arring04,Etg09,Al09}. 

In this experiment, carried out at Jefferson Lab in Hall C, a longitudinally polarized 
electron beam (82-86\% polarization) was scattered elastically off a 20 cm liquid hydrogen
target at $Q^2 = 2.5$ GeV$^2$. Electrons were detected by a 1744 channel lead-glass electromagnetic calorimeter 
(BigCal), which measured their coordinates and energy. Overlapping analog sums of up to 64
channels were used to form the BigCal trigger with a threshold of about half the 
elastic electron energy. Coincident protons were detected in the High Momentum Spectrometer (HMS) \cite{HornLongPaper}. The HMS trigger was
formed from a coincidence between a scintillator plane located behind
the drift chambers and an additional paddle placed in front of the
drift chambers. The polarization of scattered protons, after undergoing spin 
precession in the HMS magnets, was measured by the Focal Plane Polarimeter 
(FPP), which consists of an assembly of two 55 cm thick CH$_2$ analyzer blocks, 
each followed by a pair of drift chambers to track re-scattered protons with an angular
resolution of approximately 1 mrad. Elastic event selection was performed offline 
in the same way as explained in \cite{Puckett10}, resulting in a very small inelastic 
contamination for all three kinematics; at $\epsilon=0.15$, 
where it is the highest, the background fraction is $0.7\%$.

The scattered proton polarization was obtained from the angular
distribution of protons scattered in the analyzer blocks of
the FPP. The polar and azimuthal scattering angles ($\vartheta,
\varphi$) of single-track events in the FPP chambers were calculated relative to the
incident track defined by the HMS drift chambers.
The difference and the sum of the azimuthal angular distributions for
positive and negative beam helicities give the physical
(helicity-dependent) and instrumental or
false (helicity-independent) asymmetries at the focal plane, respectively. 

Since the proton polarization components are measured at the focal plane,
knowledge of the spin transport matrix of the HMS is needed to obtain
$P_t$ and $P_\ell$ at the target. The differential-algebra based modeling
program COSY \cite{COSY} was used to calculate the spin-transport matrix elements
for each event from a detailed layout of the HMS magnetic elements. The
quantities $P_{e}A_{y}P_{t}$ and $P_{e}A_{y}P_{\ell}$ were extracted using the maximum-likelihood
method described in \cite{Gayou02,Puckett10}, with $A_{y}$ the analyzing power of
$\vec{p}+CH_{2} \rightarrow \text{one charged particle} + X$ scattering. 
Their ratio gives $P_{t}/P_{\ell}$ independent of $A_{y}$ and $P_{e}$.

\begin{figure}[h]
  \begin{center}
    \includegraphics[scale=0.46, trim = 15mm 81mm 13mm 76mm, clip ]{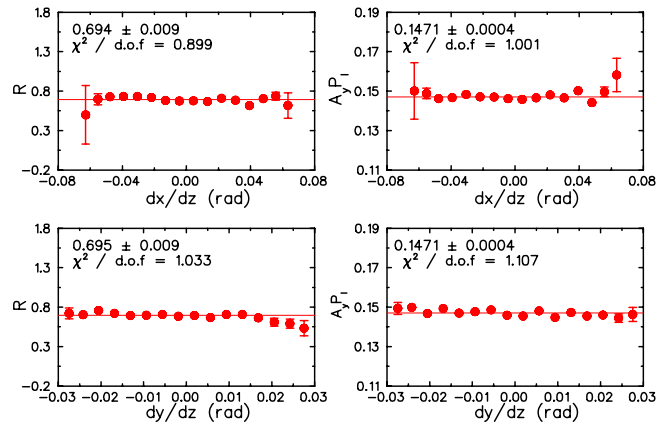}
  \end{center}
  \caption{\label{ayplprl} $R$ (left column) and $A_{y}P_{\ell}$ (right column) versus the
  dispersive $dx/dz$ (vertical) and non-dispersive $dy/dz$ (horizontal) slopes  for the low-energy setting using the COSY model.
 }
\end{figure}

As an example of the quality of the data, Fig.\ref{ayplprl} shows $R$ and $A_{y}P_{\ell}$ as a function of the vertical  
$(dx/dz)$ and horizontal $(dy/dz)$ slopes of the scattered proton trajectory relative to the HMS optical axis. 
Owing to the small acceptance of the HMS in both $\epsilon$ and $Q^2$ for all three kinematics, $R$ and $A_{y}P_{\ell}$ 
are constant across the acceptance to a very good approximation. The absence of anomalous dependence of the 
extracted $R$ and $A_{y}P_{\ell}$ on the reconstructed kinematics is thus an important test of the accuracy of the 
field description in the COSY calculations. In each panel of Fig. \ref{ayplprl}, the data are integrated over the full 
acceptance of all other variables. The horizontal line shows the one-parameter fit to the extracted data. 
In all panels, the $\chi^2$ per degree of freedom is close to one, indicating the excellent quality of the 
precession calculation.
\begin{table*}
  \caption{\label{resultstable} Kinematic table with the average quantities: the beam energy $E_{e}$, the momentum transfer squared $Q^{2}$, the 
  electron scattering angle $\theta_{e}$, and the kinematical parameter $\epsilon$. Both the ratio $R$ and longitudinal polarization $P_{\ell}$ 
  divided by Born approximation $P^{Born}_{\ell}$ are given with statistical (stat.), total systematic (tot.) and point-to-point (p.t.p.) uncertainties 
  relative to the highest $\epsilon$ point for $R$ and to the lowest $\epsilon$ kinematic for $P_{\ell}/P^{Born}_{\ell}$.}
    \begin{ruledtabular}
     \begin{tabular}{cccccccc}
       $<E_{e}>$(GeV) & $<Q^{2}>$(GeV$^2$) & $<\theta_e>(^\circ)$ & $\epsilon$ & $R \pm stat. \pm p.t.p.$	 & $ tot.$   & $P_{\ell} / P^{Born}_{\ell} \pm stat. \pm p.t.p.$ & $tot.$\\ \hline
       1.87 & 2.493 & 104.0 & 0.152 $\pm^{0.025}_{0.030} $& $0.696 \pm 0.009     \pm 0.006$	 & $ 0.013$   & $\quad- \quad\quad\quad\quad  - \quad\quad\quad\quad  -\quad$ & $-$\\ 
       2.84 & 2.490 & 44.6 & 0.635 $\pm^{0.013}_{0.017} $& $0.688 \pm 0.011      \pm 0.001$	 & $ 0.009$   & $1.007  \quad \pm 0.005 \quad \pm 0.005$ & $0.010$\\
       3.63 & 2.490 & 31.7 & 0.785 $\pm^{0.008}_{0.010} $& $0.692 \pm 0.011      \pm 0.000$	 & $ 0.009$   & $1.023  \quad \pm 0.006 \quad \pm 0.005$ & $0.011$\\ 
     \end{tabular}
    \end{ruledtabular} 
\end{table*}

\begin{figure}[h]
  \begin{center}
    \includegraphics[width=.52\textwidth, trim = 14mm 42mm 15mm 23mm, clip ]{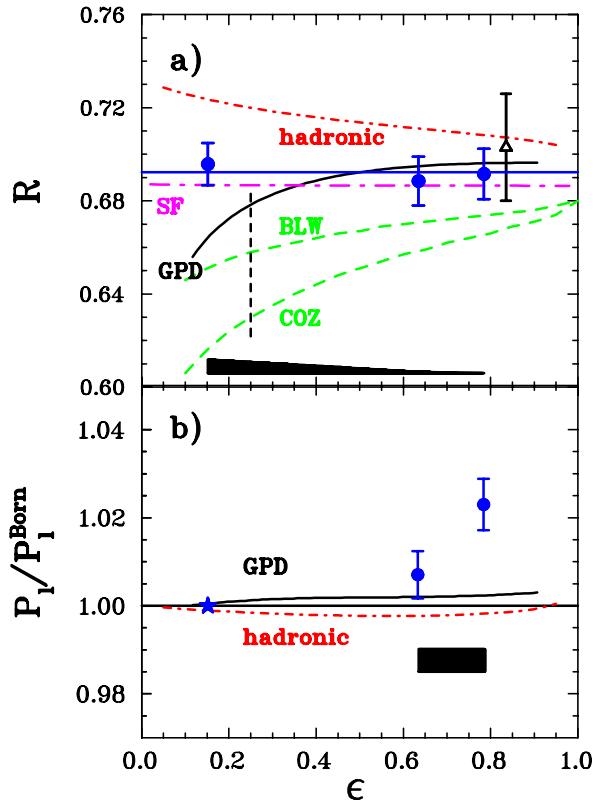}
   \end{center}
  \caption{a) \label{ratiopl}$R$ as a function of $\epsilon$ with statistical uncertainties, filled circles from this experiment and open triangle 
  from \cite{Punjabi05}. The theoretical predictions are from: \cite{Blund05} (hadronic), 
  \cite{Afa05} (GPD), \cite{Kivel09} (COZ and BLW) and \cite{Byst07} (SF) offset for clarity by -0.006 with respect to the fit. The one parameter 
  fit result is: $R=0.6923 \pm 0.0058$. b) $P_{\ell}/P^{Born}_{\ell}$ as a function of $\epsilon$. The point-to-point systematic uncertainties, 
  shown with a band in both panels, are relative to the largest $\epsilon$ kinematic in a) and relative to the smallest $\epsilon$ kinematic in b).
  The star indicates the $\epsilon$ value at which the analyzing power is determined.}
\end{figure}

The main results of this experiment are given in Table \ref{resultstable} and shown in Fig. \ref{ratiopl}. 
Figure \ref{ratiopl}a displays $R$ as a function of $\epsilon$ with selected theoretical estimates.
The data do not show any evidence of an epsilon dependence of $R$ at $Q^2=2.5$ GeV$^2$. 
Both statistical and point-to-point systematic uncertainties (relative to
the largest $\epsilon$ kinematic) are shown in the figure.
The \emph{total} systematic uncertainties in $R$ are shown in Table \ref{resultstable}. 
For a given data point, the point-to-point systematics are obtained as the quadrature sum over the differences between each of the systematic
contribution and the corresponding one for a reference kinematic. Because the dominant sources of systematic uncertainty 
described below affect $R$ for all three kinematics in a strongly correlated fashion, the systematic 
uncertainty on the relative variation of $R$ as a function of $\epsilon$, characterized by the 
point-to-point uncertainties, is very small.

Another sensitive probe of two-photon effects is taking the ratio of the 
measured $P_{\ell}$ to $P^{Born}_{\ell}$, where $P^{Born}_{\ell}$ is $P_{\ell}$ calculated in the Born 
approximation. In the limit $\epsilon \rightarrow 0$, angular momentum conservation requires
$P_{\ell} \rightarrow 1$, independent of $R$ (see Eq. \ref{eq:twogammapol}); for our measurement at $\epsilon=.15$,
$P_{\ell}$ varies by only 1.4\% (relative) \cite{pent08} for $R$ between 0 and 1. Therefore, the
measured value of $hA_{y}P_{\ell}$ at $\epsilon=.15$ determines $\bar{A}_{y}=0.15079 \pm 0.00038$ (specific to
this polarimeter), corresponding to a relative uncertainty of 0.25\%, included in the
statistical error budget for $P_{\ell}/P^{Born}_{\ell}$. Applying the same phase space cuts
at the focal plane results in $A_{y}$ being the same for all three kinematics,
at the $10^{-3}$ level. $P^{Born}_{\ell}$ was calculated from the beam energy, the proton
momentum, and the fitted value of $R$ from this experiment, with the errors
in each quantity accounted for in the total systematic error in $P_{\ell}/P^{Born}_{\ell}$.
In Fig. \ref{ratiopl}b, the ratio $P_{\ell}/P^{Born}_{\ell}$ is plotted versus $\epsilon$. The results
show an enhancement at large epsilon of $2.3\pm0.6\%$ relative to the Born
approximation.

The number of sources of systematic uncertainty is drastically reduced by the fact that the
beam helicity and the polarimeter analyzing power cancel out exactly in the ratio
of polarization components. Consequently, the
spin precession uncertainty is the dominant contribution. Since the central proton momentum was fixed across the 
three kinematics, the spin transport matrix is
identical, resulting in small point-to-point systematic
uncertainties. The error $\Delta \phi_{bend}=\pm 0.5$ mrad in the non-dispersive
bend angle, due to uncertainty in the HMS
quadrupole positions, represents the largest contribution. 
The uncertainty of $\pm 0.1 \%$ in the absolute determination of the proton momentum, which has a 
negligible effect on the precession uncertainty in $P_{t}/P_{\ell}$, leads to a relative uncertainty in the 
kinematic factor $\sqrt{\tau(1+\epsilon)/2\epsilon}$ that is roughly $1/3$ of the relative uncertainty 
in $P_{t}/P_{\ell}$ at the lowest $\epsilon$, but negligible at higher $\epsilon$. Errors in the dispersive
bend angle, the beam energy, and the scattering angle in the FPP give
smaller contributions. The inclusion of instrumental asymmetry terms obtained from
Fourier analysis of the helicity-independent asymmetries in the likelihood
function induces a negative correction to 
$R$ ($|\Delta R| \leq 0.013$) and $P_\ell/P_\ell^{Born}$ $\left(\left|\Delta P_\ell/P_\ell^{Born}\right| \le 0.004\right)$
for each $\epsilon$ value. A systematic uncertainty
equal to half the false asymmetry correction was included in the total
systematic uncertainty on $R$ and $P_{\ell}/P^{Born}_{\ell}$. 
A 1\% absolute
systematic uncertainty (0.5\% point to point) from the M\"oller measurements of the beam
polarization was added to the error  budget of $P_{\ell}/P^{Born}_{\ell}$. In order to minimize systematic differences in
the spin transport calculation among the three kinematics, cuts were
applied to the focal plane trajectories of the data
for the two larger $\epsilon$ points to match the smaller
acceptance of the point at $\epsilon=0.15$. The program MASCARAD
\cite{AfanasevRadCorr} was used to compute ``standard" radiative
corrections to $R$. Small, positive corrections $\Delta R$ of 
1.2$\times$10$^{-3}$, 1.4$\times$10$^{-4}$ and  0.7$\times$10$^{-4}$ were found for $\epsilon = 0.15,$ $0.63$, and $0.78$, respectively.
The corrections to $P_{\ell}/P^{Born}_{\ell}$ were found to be even smaller.
The results shown in Table \ref{resultstable} do not include these corrections.

The theoretical curves shown in Figure \ref{ratiopl}a make widely varying predictions for the epsilon dependence of $R$.
The hadronic model of Blunden \textit{et al.} \cite{Blund05}, where all the proton
intermediate states are taken into account via a complete calculation of the loop integral using 4-point 
Passarino-Veltman functions \cite{PassaVelt79}, shows a significant positive TPEX
contribution at small $\epsilon$. The inclusion of higher resonances
makes almost no difference \cite{Kond05}. On the contrary, the
partonic model of Afanasev \textit{et al.} \cite{Afa05}, where the TPEX takes
place in a hard scattering of the electron by quarks which are embedded in the nucleon through the GPDs, predicts
a negative TPEX contribution. A pQCD calculation of Kivel
and Vanderhaeghen \cite{Kivel09}, which uses two different light front proton distribution amplitude parametrizations, one from 
Chernyak \textit{et al.} (COZ) \cite{COZ89} and the other one from Braun \textit{et al.} (BLW) \cite{BLW06}, presents a behaviour 
similar to the partonic model. The limit of applicability of the GPD and pQCD models is shown by the vertical dashed line on the
figure. The electron structure function (SF) based model developed by Bystritskiy \textit{et al.} \cite{Byst07}, which takes into account
all high order radiative corrections in the leading logarithm approximation, does not predict 
any measurable $\epsilon$ dependence of $R$. The GPD, hadronic and pQCD models, while in good agreement with 
the available cross section data, predict a deviation of $R$ at small $\epsilon$ due to modification in $P_{t}$; 
not seen in the results presented here. Refering to Eq.(\ref{eq:twogammapol}), $R$ is directly proportional 
to the Born value $G_{E}/G_{M}$, so all the theory predictions, which use a $G_{E}/G_{M}$ value from \cite{Jones00,Punjabi05,Gayou02}, 
can be renormalized by an overall multiplicative factor. The enhancement seen in $P_{\ell}/P^{Born}_{\ell}$ is not predicted by any models.
The behiavior of $R$ at large $\epsilon$ implies the same deviation of $P_{t}$ from its Born value as the one observed in $P_{\ell}/P^{Born}_{\ell}$.

The high precision data presented in this letter add significant constraints on possible
solutions of Eq.(\ref{eq:twogammapol}) for the real part of the TPEX amplitudes
\cite{Gutt10}. In this experiment, no $\epsilon$ dependence was found in
$R$, suggesting that the $2\gamma$ amplitudes are small or compensate 
each other in the ratio. The study of the non-linearity of the Rosenbluth plot, a precise 
measurement of the single spin asymmetries and the determination of the $\sigma _{e^{+}}/\sigma _{e^{-}}$ ratio are essential 
to fully understand, quantify and characterize the two-photon-exchange mechanism in electron proton scattering.

The GEp2$\gamma$ collaboration thanks the Hall C technical staff and the Jefferson Lab Accelerator Division
for their outstanding support during the experiment. This work was supported in part by the U.S. Department of Energy,
the U.S. National Science Foundation, the Italian Institute for Nuclear research, the French Commissariat \`{a}
l'Energie Atomique (CEA), the Centre National de la Recherche Scientifique (CNRS), and the Natural Sciences and
Engineering Research Council of Canada. This work is supported by DOE contract DE-AC05-06OR23177, under
which Jefferson Science Associates, LLC, operates the Thomas Jefferson National Accelerator Facility.
\bibliography{main}

\end{document}